\documentstyle[preprint,prb,aps]{revtex}
\tightenlines

\begin{document}


\draft
\title{Quantum Dot Self-assembly in Growth of Strained-Layer Thin Films: a Kinetic Monte-Carlo Study}
\author{K. E. Khor and S. Das Sarma} 
\address{Department of Physics,
University of Maryland, College Park, Maryland 20742-4111}
\date{\today}
\maketitle

\begin{abstract}

We  use Monte-Carlo simulations to study island 
formation in the growth of thin semiconducting films
 deposited on lattice-mismatched substrates.
It is known that islands nucleate with critical nuclei of about
 one atom and grow two dimensionally until they reach a critical size 
s$_c$, when it is favorable for the islands to become three dimensional.
 We investigate the mechanism for this transition from two-dimensional(2D)
to three-dimensional(3D) growth. Atoms at the edge of 2D islands with
 the critical size s$_{c}$ become mobile as a result of strain and are
promoted to the next level. Edge atoms of the resulting island 
remain highly strained 
and are promoted to the higher layers in quick succession.
This process of depletion is rapid and occurs at a sharply defined
island size. We discuss why this leads to
 the uniformity
 seen in self-assembled quantum dots in highly mismatched heteroepitaxy
\end{abstract}
\narrowtext
\section*{I. Introduction}
There has been considerable attention in recent years on the
nature of the formation of three dimensional (3D) islands called
quantum dots (QD) {\cite{leo,moi,ger,grun} during the growth 
of strained-layer superlattices. 
For Ge grown on Si(001), for example, 
the nature of islands seen have
been characterized by Mo {\it et al} \cite{mo}.
`Hut' clusters are the first type of islands to appear with
well defined (105) facets, tilted at 11.3$^{\circ}$ to the surface,
then there is a transition to larger islands with (11n)-like
faces\cite{han} and finally to even larger but dislocated islands.
It is possible to bypass the hut cluster stage
by growing at slightly higher temperatures \cite{mo,sop}.
Another much studied system is the growth 
of InAs on GaAs(001) (mismatch $\sim 7\%$); here the particular
interest is in the  
 uniformity in the size of islands formed \cite{leo,moi}.
 This uniformity, with
dispersions of 10\% in height and 7\% in diameter of the
islands at the initial stages of formation, decreases\cite{leo}
with coverage $\theta$.
There seems to be a distinct coverage $\theta_{c}$ 
( = 1.5 monolayers (ML)\cite{leo},
1.75ML\cite{moi}, 1.7ML\cite{ger} ) at which the transition 
from two dimensional (2D) to three dimensional (3D) growth 
occurs for the InAs/GaAs system. This critical thickness
transition is slightly dependent on growth conditions; the work of
Gerard {\it et al} \cite{ger} shows that 
by substantially increasing the deposition rate, for example,
it is possible to shift it from 1.7ML to 1.95ML. There
is evidence that the material to build an island comes mainly
by depleting its immediate environment: the thickness of the 
InAs layer before islanding occurs, which is between one and two
MLs, is reduced to one ML in the immediate region surrounding the
island \cite{leo,moi,ger}. This suggests that the critical layer
thickness for InAs/GaAs(001), beyond which it is energetically
 favorable to form islands, is actually one ML and that
the extra thickness before islanding may be due to the presence
of a barrier at the 2D to 3D transition. There are signs that some
depletion is also present in the growth of Ge on Si \cite{shk,kas}.
It is known that island shapes and sizes can depend on growth conditions
\cite{schu,mol,ruv}, so that kinetic effects are important. Under much
higher deposition rates and lower growth temperatures than those used
by Leonard {\it et al} \cite{leo} and Moisson {\it et al} \cite{moi},
Ruvimov {\it et al} \cite{grun,ruv} found that
islands also exhibit size uniformity ($< 20\%$);however, 
 island size increases
with coverage \cite{grun,ruv}.
While Moisson {\it et al} observed (104) and
(110) facets on the islands, Grundmann {\it et al}\cite{grun} and
 Ruvimov {\it et al} \cite{ruv} saw only (110) facets;  
Moll {\it et al} \cite{mol} showed that the equilibrium shape of an
InAs island involves (111), $(\bar{1}\bar{1}\bar{1})$ and (110)
facets in proportions which change with the size of the island. 

Kinetic effects clearly change the shape and size of islands, and may
even affect the critical thickness for the 2D-3D transition; however,
the uniformity of islands seems to be robust for the highly strained
InAs/GaAs system. 
We first look at the experimental results of islanding in InAs/GaAs
systems because of the availability of data at small increments
of coverage $\theta$ \cite{leo,moi,ger}. There are a number of
observations that need to be discussed. The first is the narrow 
distribution in width and height of the 3D islands. The second
is the existence of a sharp (possibly first order) 2D-3D transition 
at a critical coverage $\theta_{c}$ \cite{leo}. There is also
the phenomenon of fast depletion (of the order of seconds \cite{ger}
), where a 3D island is created quickly (compared to a deposition rate
of .01ML/sec \cite{leo}) largely
out of the atoms from the its 2D environment. Finally, it is also
seen that under conditions of slow deposition 3D islands
remain essentially constant in size over a coverage interval 
of $\Delta\theta \sim 0.4$. Note that these results are affected
by growth conditions. Under high deposition rates
(compared to diffusion rates) which is possible at low temperatures, the sharp
island size distribution may disappear, see, for example,
growth of Ge on Si(001) \cite{kas,schu,med}, where the  lattice mismatch
is smaller ($\sim$ 4\%). For the Ge/Si system, where the strain is much
less than that in the InAs/GaAs system, depletion seems to take a time
of the order of minutes at 550$ ^{\circ}$C  \cite{shk,kas,med} and recent growth
experiments were carried out at typical deposition rates of a few ML's/sec
\cite{kas}. Under these growth conditions, even the sharp 2D-3D
transition may disappear\cite{pol}.

In this study then we focus on the early stages of growth for thin films 
which grow in the 
Stranski-Krastanov (SK) mode.  We study growth under conditions where
diffusion is fast compared to deposition, so that effects due to the process
of depletion can be distinguished from those due to deposition.
In a previous work \cite{kho}, we investigated the energetics of the 
2D to 3D transition 
in detail by means of molecular dynamics simulation,  using an
empirical potential that has been appropriately tuned \cite{kho}.
We argued that the 2D to 3D transition
occurred when 2D islands had grown much larger than the
size $s_{o}$, when 3D islands \underline{first} become
 energetically favorable; this
effectively is a barrier, which once scaled by a 2D island,
allows  it to reorganize itself into a 3D shape, with an 
\underline{immediate}
gain of energy. This gain, which is more pronounced for the
highly mismatched InAs/GaAs system than for the Ge/Si, can
be quite substantial, about 5 - 10 meV/atom for the former. We
feel that this is the underlying factor for the uniformity
of sizes of islands seen in this system. Priester {\it et al}
\cite{pri} have attempted to provide an explanation for
the uniformity of the 3D islands, but have not taken into account the
factor of the barrier, which should affect their considerations.

It is known for the growth of Si/Si(001) 
\cite{kho2,lag,ven} that islands nucleate with a critical size of one to
three atoms and then grow two-dimensionally. This picture of nucleation
is also supported by the results of Chen and Washburn\cite{chen},
who used a critical nucleus of $i=1$ in the scaling function $\Phi(N/\bar{N}
)$\cite{ama} in fitting the island density results
 of Leonard {\it al}\cite{leo}. Island 
nucleation of Ge on Si(001) should be similar (in both cases the
dimer is the stable nucleus). We suggest that the 2D--3D transition
picture is the following: 2D islands nucleate
with critical nuclei of about one atom and grow 
two-dimensionally until a critical size $s_{c}$ when
strain makes it favorable for there to be a transition
to 3D growth. This size $s_{c}$ is quite large, roughly 
a few hundred angstroms.
There is direct experimental evidence
for this picture of growth. 
Mo and Lagally observe\cite{mo1},
 after growth
of about 3ML  of Ge on Si at 500$ ^{\circ}$C, a growth front roughness 
of three layers over an area of 60nm  x 60nm . Gerard {\it et al}
\cite{ger} observe one layer roughness over extensive 2D areas ($\sim 2000
\AA$ ) for the growth of InAs on GaAs(001) at 520$^{\circ}$C. We stress that
$s_{c} >> s_{o}$, the size at which 3D clusters have just become energetically
favorable.
 Indeed the 2D island must reach a size comparable to that of  
 the two-layer island when the latter
 becomes energetically favorable.
Once this size barrier is reached, the transition to islands of two
or more layers in height is possible since taller islands 
are already favorable at smaller sizes. 
There is a rapid rearrangement of its atoms in order to
 achieve the shape of the optimally
 energetic (105) facetted clusters. 
  There is an immediate gain in energy
of 1-2 meV/atom for the Ge/Si system; for InAs/GaAs, we estimate
this gain, assuming that the elastic energy scales with the
square of misfit, to be 5-10 meV/atom\cite{kho1}. This latter amount
is substantial and is probably the reason for the
phenomenon of depletion seen in the highly mismatched
systems.

The above picture obtained from an energetics study is complemented by
our work here on kinetics. In this study, we approach island growth
on strained-layer superlattices, by using finite temperature
non-equilibrium Monte Carlo (MC) simulations, where diffusion rates
of adatoms depend on strain as well as the usual local bonding.
Computational time and size constraints force us to carry out our
kinetic MC simulations in 1+1 dimensions, i.e., in our MC simulations
the substrate is one dimensional and the growth is two dimensional.
We do not believe that our 1+1 dimensional simulations introduce
any qualitative complications, although it will be necessary in the future
to verify our proposed picture using the full three dimensional MC
simulations.
Our results show that under growth conditions of fast diffusion
 relative to deposition, i.e., not very low growth temperature,
 the picture obtained from energetics is
largely correct. There is a sharp 2D--3D transition which occurs
at an island size s$_{c}$ which is well beyond the critical size 
s$_{o}$ at which the 3D islands first
become energetically favorable. Depletion is observed and narrow
3D island distributions are obtained. The average size of 3D islands
does not change with coverage. In this work, we attempt to understand
the microscopic dynamics and mechanisms underlying these results.
In the following section we describe the simulation method and the 
parameters chosen. Then we present detailed results of the simulation
in Sec. III and discuss the results in Sec. IV. We conclude in Sec. V.

\section*{II. The Simulation Model}
In our MC growth simulations (which is done in 1+1 dimensions),
  an adatom moves (under solid-on-solid
restrictions (SOS)) by hopping  randomly 
to neighboring sites at a rate that depends on its bonding. (We
obey detailed balance in our kinetic MC simulation.)  The hopping
activation energy depends on the bonding environment and the elastic
energy associated with strain The hopping
activation energy depends on the bonding environment and the elastic
energy associated with strain.
The hopping rate is given by the expression,
\begin{equation}
R_{n}=R_{o}exp^{-\frac{E}{k_{b}T}},
\label{equ:e1}
\end{equation}
where $R_{o}=d'kT/h$ is a characteristic vibrational frequency
and $d'=1$ is the substrate dimension. The activation energy 
$E=E_{bond}-E_{strain}$, with $E_{bond}$ being determined by the
number of nearest neighbors(nn) and next nearest neighbors(nnn).
The elastic energy is given by harmonic interactions between an
atom and its nn and nnn neighbors, using spring constants k.
Following Orr {\it et al}\cite{orr}, we obtain $E_{strain}$ for 
a particular site by taking the
difference in elastic energies of the system when the site is
unoccupied and when the site is occupied.
This energy is calculated by allowing atoms in a 5(height)x7(width)
cell centered at the site first to equilibrate under molecular
dynamics simulation and then to relax to its minimum energy configuration
by means of the method of steepest descent. Every 100 time steps
or so the entire system is allowed to relax globally to avoid
any local strain accumulation. $E_{bond}$ is chosen in the
following way, 
\begin{equation}
    E_{bond} = \left\{ \begin{array}{ll}
               E_{o}=(0.7NN+0.2NNN)eV,& \mbox{if NN $\leq2$}\\
               E_{1}=4.0eV,& \mbox{ if NN=3}\\
               E_{2}=1.45eV,& \mbox{ steps of height $\geq2$}
              \end{array}
              \right. 
\label{equ:e2}
\end{equation}
where NN is the number of nn's and NNN is the number of nnn's. $E_{o}$
applies to single adatoms or atoms at step edges, except when step
heights are two layers or greater. Then $E_{2}$, a reduced barrier height, 
is applied to the surface atoms on top of
these steps, so that inclined (11) island facets
are favored over vertical ones. 
 $E_{1}$ is the barrier for the rest of the surface atoms which have 
three nn's. It is chosen a little higher than that given by bond counting
to eliminate intrasubstrate breakaway (especially at the foot of islands)
and therefore to avoid substrate roughening,
which is not seen experimentally\cite{ger,mo}.
For simplicity we have also used the same barrier
for midisland surface atoms; results are not different from those
using bond counting for these atoms.
The parameters have been chosen so that
diffusion will dominate over deposition, for example, a single adatom
will diffuse a distance of approximately 600 unit cells for each deposition
event at 750K. This is about 50-100 times the width
of the islands that form. Using diffusion rates from Mo {\it et al}
and others\cite{mo,sri} and experimental deposition rates
and island sizes\cite{moi,leo}, we get comparable results
of the ratio diffusion distance/island size$\sim 100$.    
We choose the spring constant $k=200eV \sim 200$ times the
diffusion barrier for a single adatom\cite{orr}, and a deposition
rate of 0.01-0.2 MLs/sec. We carried out simulations for
strained-layer lattices with misfits of $0-7\%$, at temperatures
of 700 to 800K. We start with systems at thicknesses
of 11MLs, with the three top layers at the larger lattice
constant. System sizes vary from 500 to 8000 cells.
 At zero strain, growth was layer
by layer as would be expected under the above conditions of
fast diffusion -- there is no kinetic roughening at this
"high temperature" growth in the absence of strain.

\section*{III. Results}
We report on two preliminary studies that will help in
understanding the final results.
First we carry out simulations for the unstrained system, varying
the diffusion barrier for atoms at the ends of islands, $E_{end}$, from 1.3
to 1.8eV, for temperatures T, from 700-800K. and deposition rates
0.1-0.2 MLs/sec and over coverages of $\theta$ from .5 to 0.8. 
In Table \ref{tab1}, we display the results for two growth temperatures,
750 and 800K, with a deposition rate of 0.2MLs/sec, a system
size of $10^{4}$ cells and a coverage $\theta \sim 0.6$. We calculate 
a roughness index (R.I.) as
the percentage of sites in islands, which have heights $ > 1$,
i.e., R.I. is a rough measure of the deviation from "two-dimensionality"
(one-dimensionality in our simulations) in the islands.
For $E_{end} \geq 1.5eV$ growth is smooth, islands are flat (very small R.I.),
but growth is distinctly rough for $E_{end} \leq 1.4eV$,
there being much larger proportions of islands with 2 or more layers in
height. Clearly the transition from smooth to rough growth is
sharp.

Next we look at island-end energies $E_{end}$
for some island configurations when elastic interactions
are included. Specifically we carry out calculations for
misfits of 5\%. In Fig.~\ref{fig1}, we plot
$E_{end}$ against island volume (number of atoms), for 
seven island configurations, comprising (a) 1-level islands (h=1),
(b) 2-level islands (with 1 atom (h=2a) and 2-atoms (h=2b)
on the second level, (c) 3 level islands with 1 (h=3a) and 2
atoms (h=3b) on the third level and 3 and 4 atoms respectively
on the second and (d) 4 and 5 level islands each with 1 atom
on the top level (h=4,5 respectively) and the same shape
as islands in (c). Island volumes are varied by changing
the length of level 1 of the islands, while keeping
upper configurations fixed. If we take $E_{end} < 1.5$eV as the
condition for rough growth, then islands with volumes $>$ 15
will have end atoms with diffusion barriers $< $ 1.5eV for all
the consequtive configurations 1,2,3 and higher levels. The
following picture of 3D islanding is suggested: 2D islands
grow two dimensionally until a certain size when end atoms
are promoted to the second level; this process becomes more
rapid as it proceeds because $E_{end}$ increases with
the number of atoms on the second level (while island
volume is kept constant). This process then continues
in the same fashion with the subsequent promotion of atoms
to the third and higher levels. This, we believe, is the
mechanism for the phenomenon of depletion seen experimentally
\cite{leo,moi,ger}.

We now present results of our full kinetic MC simulation done on systems
of substrate sizes L=2000, 4000 and 8000 cells. The observations we report
below are true of all these sizes and so are not affected by
finite size effects. For these simulations, we also consider
the effect of a strain enhancing factor $F_{end} = 1.0,1.2$ and
1.5 on the first level end atoms of islands. It is known that
there is tremendous strain at the foot of islands\cite{har,chr}.
Our results are not particularly sensitive to variations in this strain,
aside from making islands a little smaller as $F_{end}$ is increased. 
In Fig. \ref{fig2} , we follow the development of a single island
over a growth period of about 0.2ML ($< 3$secs.). Figs. 2a,b,c show
a 1-level(2D) island of volume 19 atoms being folded up into a
2-level island in 0.5 secs. The material for this 2-level island
(volume=20) comes almost completely from the original 1-level island.    
In the rest of Figs. 2d,e,f and g, we see similarly rapid buildups 
of the third and fourth levels after a brief waiting period. The 
whole process starting from Fig. 2a to Fig. 2g takes less than
3 secs. The bulk of the material ( $\sim 80\%$ for the 3-level island
and $\sim 65\%$ for the 4-level island) for the formation of the
3D island comes from the original 2D island (compare with the
experimental results of the three groups above\cite{leo,moi,ger}).
Fig. \ref{fig2}  shows a typical 2D-3D transition sequence for islands
in our simulation. It clearly illustrates the process of depletion
seen experimentally. (Note that in our simulations what we
refer to above as 2D and 3D are really 1D and 2D respectively
since we are using 1+1 dimensional simulation).

In Fig. \ref{fig3} we display width and height distributions
of islands for a range of coverages $\theta = 0.393 - 0.87$. There
is uniformity in the island size distributions which are
sharply clustered around the mean width or height, each with
a half-width of $\sim 1$ cell. Furthermore, while island density
increases with coverage, the average island size remains
essentially  constant. In Table \ref{tab2}, we show the
average volumes at which islands undergo transitions from
the first to the second levels, from the second to the third,
and from the third to the fourth level. The root mean square deviation
is 2 atoms in each case, showing that transitions occur at sharp
distinct sizes.     

We plot in Fig. \ref{fig4} the total number of islands with
3 or more levels as a function of coverage $\theta$ for systems
of size L=4000. The results are the same for systems of other sizes
(L=2000 and 8000) when appropriately normalized. We see that  
island density is zero until a certain coverage $\theta_{c}$
is reached, when the density increases rapidly. 
Leonard {\it et al}
\cite{leo} observed this experimentally and fitted the
island density $\rho_{isl}$ with the function
$\rho_{isl} = \rho_{o} (\theta - \theta_{c})^{\alpha}$.  
They obtained a value of $\alpha =1.76 $ while we get $\alpha =1.34 $.
The difference in the value of $\alpha$ could be due to
 our using a 1+1 dimensional simulation. 
We arrive at similar conclusions if we look at islands with
2 or more levels instead of the $\geq$3 levels we have chosen above.

In Table \ref{tab3}, we show the energies of islands of various
configurations comprising 2,3 and 4 levels, relative to the
energies of their corresponding 1-level configuration at the
same volume. We see that the first energetically favorable
2-level island is the one whose volume is 8 atoms with a configuration
of 3 atoms on the second level and 5 on the first. 3-level
islands become favorable at a volume of about 12 atoms but
for this volume the 2-level configuration has the best energetics.
3 and 4-level islands are energetically optimal at volumes of
15 atoms and 24-28 atoms respectively.  These figures can be 
compared to the transition volumes of Table \ref{tab2}. There
clearly is a correspondence between energetics and kinetics.
However, one interesting point emerges, although a 2-level
island becomes energetically favorable at a volume of 8 atoms,
kinetically the transition occurs at a volume well beyond
that (around 19 atoms). Energetics sets the lower size limit for
the beginning of depletion, but it is kinetics that determines
the actual point. This is the s$_{c} >>$ s$_{o}$
kinetically driven scenario we discussed
before. Note also that because of the small size
of the islands we encounter here, it is the 2-level island
that first becomes energetically favorable before the 
taller islands; we expect the situation to be reversed when mean
island sizes are larger as surface energies become less
significant - this would be the case with the sizes actually
 seen experimentally. This aspect of physics is not appropriately
captured in our small system 1+1 dimensional simulations.  

\section*{IV. Discussion}
The experimental results of islanding in InAs/GaAs
systems of a number of groups are shown in Table \ref{tab3}. 
The first four groups observed uniformity in the size
distributions of the islands, in particular, Leonard {\it et al}
reported  dispersions of 10\% in 
height and 7\% in diameter at the first appearance of the islands,
at $\theta \sim \theta_{c}$; with further deposition,
this uniformity is reduced, island density increases but
sizes remain essentially the same \cite{leo,sol}. The first three
groups concluded that there is depletion-like behavior. 
Leonard {\it et al}
show that more than 80\% of the atoms to form an island comes
from its environment, rather than from  additional deposition. 
Gerard {\it et al} display an atomic force micrograph ( Fig. 3
in ref.\cite{ger}) of the depletion zone around an island, whose
size is $\sim 1000\AA$. They also show that the timescale of
this mass movement to form an island is from 2 to 10 seconds.
This phenomenon of depletion is clearly consistent with the results
of our simulation. It takes a few seconds in a highly
mismatched system,  but is much longer, $\sim$minutes,
in the Ge/Si system; so in this system it is probably
 masked by the deposition rates
used and only two groups have reported seeing it in this 
system \cite{shk,kas}. From an energetics perspective, beyond the
critical coverage $\theta_{c}$, as deposition continues, there
is much more energy to be gained for the new material to
 create new 3D islands than to grow existing ones.
 So there is an increase of island density but little size gain. 
 Our simulation shows that the process of depletion is driven by two factors.
First 2D islands are grown well past the size $s_{o}$ at which
3D islands become energetically favorable. At a distinct critical
size $s_{c}$ determined by kinetics, atoms at the 2D island ends are,
as a result of strain, easily promoted up to the next level. Secondly,
strain continues to be adequate to keep island-end atoms mobile even
with little further island growth, so that the next higher levels
are formed quickly, also at distinct sizes. The process of depletion
lasting seconds only, then, is largely the pulling in of existing material
to form 3D islands. It  is completed when the tallest island is formed. 
Subsequent growth of these islands is mainly by the formation of new facets.
Facet formation is generally much harder than adding atoms to the ends of
 a 2D island.
As can be seen from Fig.~\ref{fig1}, the diffusion barrier for an atom
at the end of a 2D island of size 10-15 atoms is $\sim 1.5$eV while
it is $\sim 1.38-1.4$eV at the bottom edge of a 3-4 level island. 
This difference translates into a substantial difference in mobility,
as we have seen above, so that as long as 2D islands are present, their
growth is strongly favored over that of 3D islands.  The uniformity
of islands at coverage of $\theta_{c}$ is due to depletion occurring
at distinct sizes. The continuing size uniformity coupled with
constant mean 3D island size while 3D island density increases, as deposition
proceeds, especially for $\theta -\theta_{c} \leq 0.5$ \cite{leo},
is due to the preferred growth of 2D over that of 3D islands.
Chen and Washburn\cite{chen} obtained results of continuous increase
in the size of 3D islands (see  their Fig. 5\cite{chen}) with the rate
being the largest at the smallest size. Clearly this can only apply
after most of the 2D islands have disappeared.

As we have noted before, there should be a correlation between 
kinetics and energetics. Our kinetic MC simulation shows that the depletion
process begins once strain enables 2D island edge atoms to be mobile enough
to be promoted to the upper levels. But this process must be also 
favored by the energetics; we expect the corresponding
 3D island to be energetically
more favorable than the 2D island. For the systems studied here we
have observed this correspondence. In a previous paper we studied the
energetics of (10n) facetted Ge `hut' clusters on Si
substrate\cite{kho}, using the atomic configuration of (105)
side facets suggested by Mo {\it et al}\cite{mo}. The general
conclusion was that taller (105) facetted islands become energetically
favorable at smaller sizes than islands with (10n) facets for n$\geq$7.
(103) facetted islands are excluded because these faces require
costly double steps so that they are not observed
experimentally. (105) facetted hut clusters become favorable
only when they have at minimum, heights of 12 layers\cite{kho}
(at size s$_{o}$, say).
Islands with lower aspect ratios have to reach greater sizes
to become energetically favorable. For systems which are growing
with growth front roughness of 1-2MLs\cite{mo}, it is then
necessary for 2D islands to grow well beyond the size s$_{o}$
before the transition to a 3D shape can begin. This size s$'_{c}$ may
be comparable to the size at which a {\bf 2-level} island first
becomes favorable. As we have noted in the simulations above, the actual
transition size, s$_{c}$ is determined by kinetics but this size
must be such that s$_{c} \geq $s$'_{c}$. So energetics sets the
lower size limit at which a 2D--3D transition can occur.
In Fig. ~\ref{fig5}, (see Khor {\it et al}\cite{kho1}) (111) facetted islands
are shown to become energetically favorable at sizes 
and heights greater than those for (105) islands at size
s$_{o}$. With increasing size, (111) facetted islands quickly become
more favorable than the (105) hut clusters. These results are
consistent with experimental observations of Hansson {\it etal}\cite
{han}
who obtained (111)-facetted islands under near equilibrium
conditions and also the results of Mo{\it et al}
where macroscopic structures were seen to be the stable ones.

Three groups, Leonard {\it et al}\cite{leo}, Moisson 
{\it et al}\cite{moi} and Gerard{\it et al}\cite{ger},  
observe the presence of a critical coverage $\theta_{c}$ below
which no 3D islands are seen. Leonard {\it et al} characterise
this transition to be like that of a first-order phase transition.  
We observe a similar transition in our simulations.
However, in contrast to the results  above
, Polimeni {\it et al}\cite{pol} report a smooth 2D--3D transition
for the growth of InAs on GaAs(001). In Table \ref{tab4}, we
compare the growth conditions for the different groups.
The growth temperature used by Polimeni {\it et al}
at $420^{\circ}$C, is substantially lower than those ($500-530^{\circ}$C)
used by the other groups. At these temperature differences, the diffusion
rate D could differ by more than an order of magnitude. Assuming a
behavior for D to be similar to that observed for Si adatoms on 
Si(001)\cite{lag,ven}, we calculate the ratio R/D, where R is the
deposition rate. This ratio is $\sim N^{3}$ (N=
 island density), for the same 
coverage\cite{lag,ven}; this tendency to nucleate islands
 should correlate with growth front roughness.
 We see from the last column of Table \ref{tab4}, 
this ratio  for Polimeni {\it et al} is about 50 times that for
 Leonard {\it et al}.   
Since higher effective deposition rates contribute to rougher growth,
it may make level to level transitions less distinct than those we
have seen in the simulations above. At some point kinetic
roughness at the growth front arising from fast (slow) 
deposition (diffusion) 
may mask the phenomenon of depletion and give rise to an
 apparently smooth 2D--3D transition.

 Solomon{\it et al}\cite{sol} have shown that
 3D island density, at fixed coverage
and temperature, is increased when  either the growth rate
R is reduced or the diffusion D is increased. (The latter is done by increasing 
 the flux V/III ratio)\cite{sol}. A related observation is made
by Mo and Lagally\cite{mo1} for Ge/Si(100) growth. When they
deposited Ge at 850K , they found the concentration of macroscopic
clusters to be higher than that at T$<$800K.
  This result is unexpected for 
 as we have seen above from nucleation theory for
regular island growth that island density 
 goes as $R^{p}/D^{p}$ where
p is positive \cite{ven}; but this applies, in
our case to  the \underline{2D} islands. Increasing deposition rate or
decreasing diffusion then increases the 2D island density and
correspondingly decreases average 2D island size at a given coverage. 
Assuming that there is an average 2D-3D island transition
size s$_{c}$ for the growth regimes of Solomon {\it et al} (this must be
the case since they observe the constant 3D island diameter throughout
 their experiments),
 this means that fewer 2D islands reach this size at that coverage. 
 The density of \underline{3D} islands then increases by 
 reducing growth rate or increasing D. We see from table \ref{tab4},
that its relative R/D ratio of 24-52 may put it in a "rough" 
 growth regime closer
to that of Polimeni {\it et al} than to that of Leonard {\it et al},
so that the existence of a sharp $\theta_{c}$ is uncertain. 

 In Table \ref{tab4} it is interesting to note that even for a
relative R/D=120, much larger than that of Polimeni {\it et al}\cite{pol},
Ruvimov {\it et al}\cite{ruv} still observed 3D island size uniformity
$<20\%$. This is true even when their observations were carried out for
coverages of $\theta$ = 2-4MLs, which is much greater than the $\theta_{c}
$ of Leonard {\it et al}. They did not specifically study if the
2D-3D transition is sharp or smooth.     

We have noted above that depletion is seen to occur on a timescale
of minutes in Ge/Si systems compared to seconds for InAs/GaAs systems.
 We should expect to see results in the former similar
to those observed for the latter if the deposition rates and
growth temperatures are appropriately scaled. Shklyaev {\it et al}
\cite{shk} have carried out growth experiments of Ge on Si(111) 
at small increments of coverage;
they used a growth rate of 0.004 bilayer(BL)/sec and a temperature
of $480^{\circ}$C ( R/D$\sim$0.8, see Table \ref{tab4}).
 They observed the growth of two types of islands
which were called large flat islands and 3D islands. The 
latter appear 'abruptly', there being a distinct jump in 3D
island density over a growth interval of 0.1 BL. Much of the
material for the formation of these islands come from the
substrate. Annealing experiments suggest that this depletion
occurs over a time period of about 10 minutes. They did not measure
island size distribution but their Fig. 1 shows 3D island images
which appear quite uniform in size. Many of the experiments for 
the growth of Ge on Si were carried out with quite high relative
R/D values, for example, R/D=7,7 and 4 for Voigtlander {\it et al}\cite{kas}  
, Medeiros-Ribeiro {\it et al}\cite{med}  
and Kastner {\it et al}\cite{kas} respectively. It is in this
growth regime that the last two groups observed rectangularly
shaped `hut' islands\cite{mo}. Voigtlander {\it et al} saw
the aspect ratio of a single island change over a coverage
interval of $\sim$1BL(20min), indicating that depletion probably
takes that long. 

As noted above, the time ($\tau)$ for depletion increases with decreasing
  lattice
mismatch $x$. In general it probably goes as $\tau \sim x^{-\eta} D^{-\gamma}$,
where $\eta$ and $\gamma$ are some positive constants. We have seen
above that depletion occurs over a large range of growth conditions
as determined by the ratio R/D. We would expect depletion to fail to
occur only when it is completely overwhelmed by deposition, that is,
when 1/R $\gg \tau$, or when $x \ll R^{1/\eta}D^{-\gamma/\eta}$.   
This must be the condition for smooth (non-islanding) growth at
low temperatures or high deposition rates. The relationship
must be applicable to the temperature-concentration phase
curve, delineating smooth from rough growth for the deposition
of Si$_{1-x}$Ge$_{x}$ on Si(001) obtained by Bean {\it et al}\cite{bea}.
In Fig. ~\ref{fig6}, (Khor {\it et al}\cite{kho1}), a replot of the experimental
data of Fig.1 of Bean {\it et al} shows a linear relationship of
ln(x) versus 1/T, (except for the point at $x=1$ where a minor
temperature change from 550 to 527$^{\circ}$C would put the point on
the line), which supports this conclusion.

\section*{V. Conclusion}
In conclusion, we find that in general, for strained heteroepitaxial
growth of semiconductors, there exists an effective kinetic barrier
for the 2D to 3D transition. Under conditions of slow deposition and fast
diffusion, islands initially grow two-dimensionally
to a size $s_{c} >> s_{o}$  well beyond the size $s_{o}$  
 at which a 3D island {\bf first}
becomes energetically favorable. At this size $s_{c}$ atoms at the edge
of the 2D island become mobile as a result of strain, and are
promoted to the next level. Promotion of atoms to the next levels occurs in
quick succession because edge atoms continue  to be highly strained  
and so remain mobile. The process of depletion is completed when the
island attains it highest aspect ratio. This size $s_{c}$ is sharply defined,
and there is a correlation with energetics. 
  This is a robust result that should apply to
a wide  range of semiconductor systems. For highly mismatched systems, it is the
underlying microscopic reason for the uniformity in the sizes of islands seen
experimentally. It is consistent with other experimental results such as
the increase in island density with coverage with no corresponding
increase in size, the phenomenon of depletion, 
the (initially unexpected), result that island
density increases with reduced growth rate or enhanced diffusion.

This work is supported by the
U.S.-ONR and NSF-MRSEC.

\begin{table}
\caption{Growth roughness of the unstrained
system as a function of island-end diffusion barrier height.}

\begin{tabular}{|ccccc|} \hline 
\multicolumn{1}{c}{T}& \multicolumn{2}{c}{750$^{\circ}$K}&
\multicolumn{2}{c}{800$^{\circ}$K} \\ \hline
$E_{end}$&$\theta$&R.I.&$\theta$&R.I.\\
0.3&0.56&29.3&0.56&41.0\\
0.4&0.63&7.4&0.65&17.6\\
0.5&0.61&1.3&0.63&3.2\\
0.6&0.64&1.1&0.62&1.3\\
\end{tabular}

\label{tab1}

\end{table}

\begin{table}
\caption{Mean volumes at which islands undergo transitions
from levels 1-2, 2-3 and 3-4.}

\begin{tabular}{|cccc|} \hline 
transitions from levels&1-2&2-3&3-4\\
island volumes&19.0 $\pm 2$&21.4 $\pm 2$&26.3 $\pm 2$\\
\end{tabular}

\label{tab2}

\end{table}

\begin{table}
\caption{Energetics of islands of various configurations
relative to the energy of the corresponding 1 level island at the
same volume.}

\begin{tabular}{|ccccc|} \hline 
\multicolumn{1}{c}{}& \multicolumn{3}{c}{island configurations}&
\multicolumn{1}{c}{} \\ \hline
volume&2-level&3-level&4-level&$\Delta E$\\
4&1,3&&&-0.58eV\\
6&2,4&&&-0.18eV \\
8&3,5&&&0.14eV\\
9&&1,3,5&&-0.24eV\\
12&5,7&&&0.72eV\\
12&&2,4,6&&0.42eV\\
14&6,8&&&0.97eV\\
15&&3,5,7&&1.06eV\\
24&11,13&&&1.79eV\\
24&&6,8,10&&2.72eV\\
24&&&3,5,7,9&2.6eV\\
28&13,15&&&1.28eV\\
28&&&4,6,8,10&3.48eV\\
\end{tabular}

\label{tab3}

\end{table}
 
\begin{table}
\caption{Growth regimes of various experimental groups}

\begin{tabular}{|l|cccccc|} \hline 
&Leonard\tablenotemark[1]&Moison\tablenotemark[2]&Gerard\tablenotemark[3]&Ruvimov\tablenotemark[4]&Polimeni\tablenotemark[5]&Solomon\tablenotemark[6]\\ \hline
Growth Temperature&530$^{\circ}$C& 500$^{\circ}$C& 520$^{\circ}$C& 480$^{\circ}$C& 420$^{\circ}$C& 500$^{\circ}$C\\
Deposition rate ML/sec&0.01&0.06&0.06&0.6&0.1&.18-.39\\
Island uniformity&24\%&40\% &&$<$20\%&&\\ 
$\theta_{c}$&1.5&1.75&1.7&-&-&-\\
Relative R/D&1&8&6&120&50&24-52\\ \hline

\end{tabular}
\tablenotemark[1]{Reference\protect\cite{leo}.}
\tablenotemark[2]{Reference\protect\cite{moi}.}
\tablenotemark[3]{Reference\protect\cite{ger}.}
\tablenotemark[4]{Reference\protect\cite{ruv,grun}.}
\tablenotemark[5]{Reference\protect\cite{pol}.}
\tablenotemark[6]{Reference\protect\cite{sol}.}

\label{tab4}

\end{table}

\begin{figure}
\caption{ 
Energy of atoms $E_{end}$ at ends of islands plotted against
island volume. The island configurations are:
(a) 1-level islands (h=1), (b) 2-level islands with 1 atom (h=2a)
and 2 atoms (h=2b) on the second level, (c) 3-level islands with
1 (h=3a) and 2 atoms (h=3b) on the third level and 3 and 4 atoms respectively
on the second level, (d) 4 and 5 level atoms with 1 atom on the top level
(h=4,5 respectively) and the same shape as in (c).
}
\label{fig1}
\end{figure}

\begin{figure}
\caption{ The process of depletion: 
A 2D island undergoing a transition to the 3D shape. (a) Shows the
2D island just before it is rapidly folded up into a 2-level
island in (b) and (c). Figs. (d), (e), (f) and (g) show rapid promotion to
the upper layers.
}
\label{fig2}
\end{figure}

\begin{figure}
\caption{ 
Height and width distributions of islands at coverages $\theta$=
0.4ML,0.67ML,0.79ML and 0.96ML. 
}
\label{fig3}
\end{figure}

\begin{figure}
\caption{
Plot of 3D island density versus coverage $\theta$ for system sizes
L=2000, 4000 and 8000. Island density is normalised to that for L=4000.
}
\label{fig4}
\end{figure}

\begin{figure}
\caption{ 
Island energies in meV/atom versus cluster size. Dashed line:
rebonded (105) islands
; light solid: (107) islands; dot-dashed line: (111) islands;
heavy solid line: two layer islands. (105),(107) and 2-layer
island curves cross the abscissa at points A,B and C respectively.
}
\label{fig5}
\end{figure}

\begin{figure}
\caption{
Replot of the film morphology curve of Bean {\it et al}
(Fig. 1 in ref.\protect\cite{bea}), 1/T, in $^{\circ}K^{-1}$x1000, versus ln(x). 
}
\label{fig6}
\end{figure}
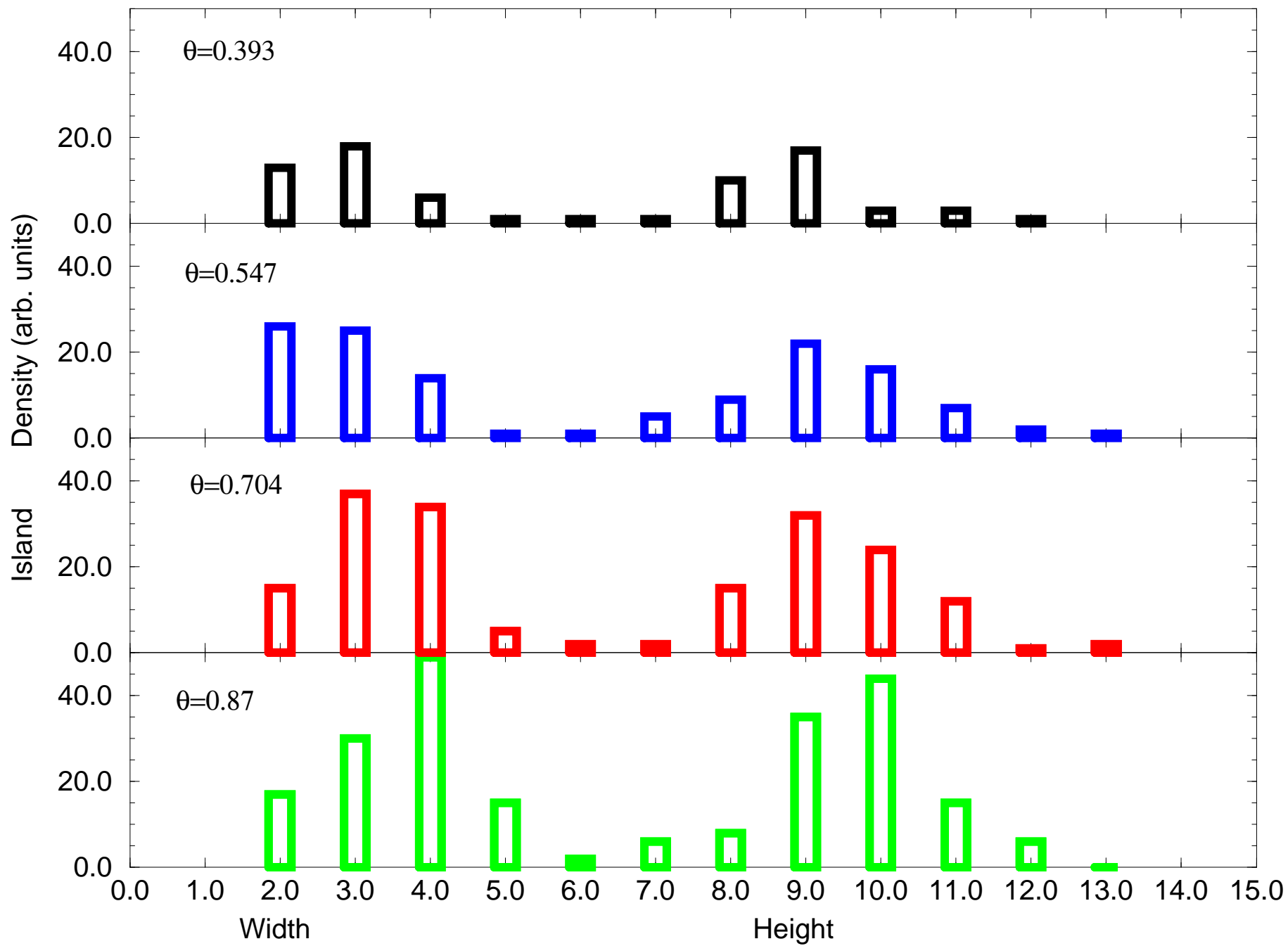
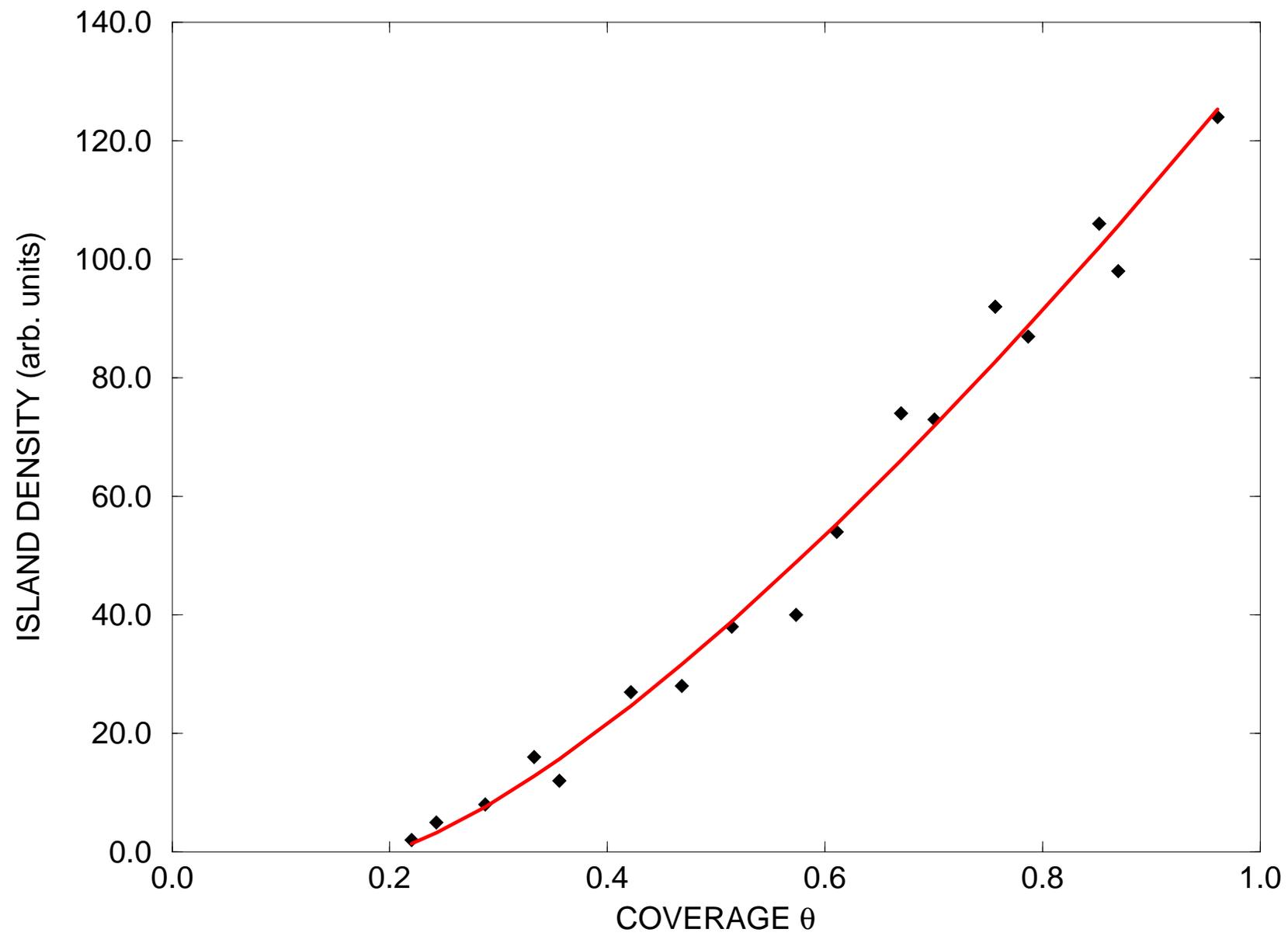
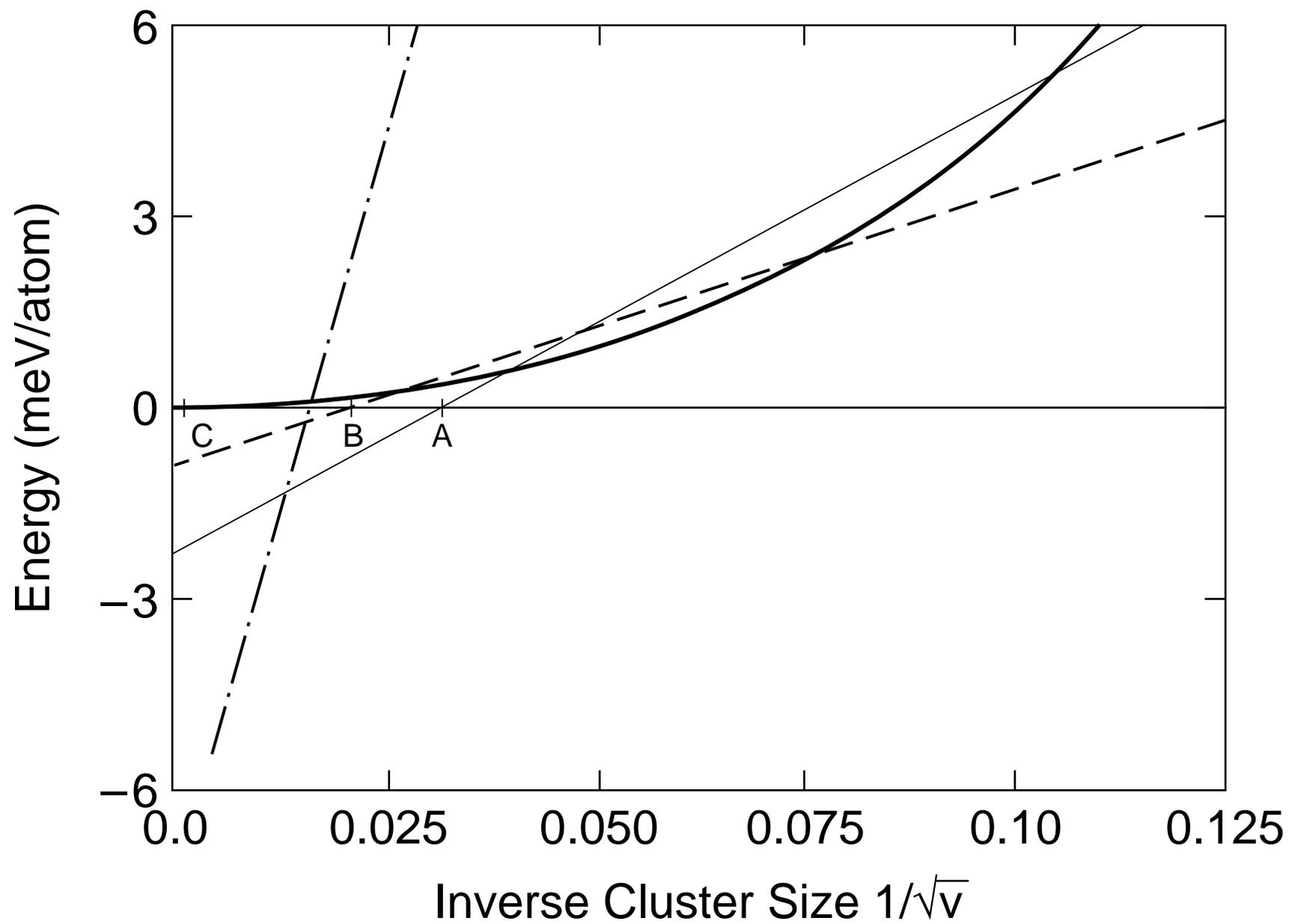
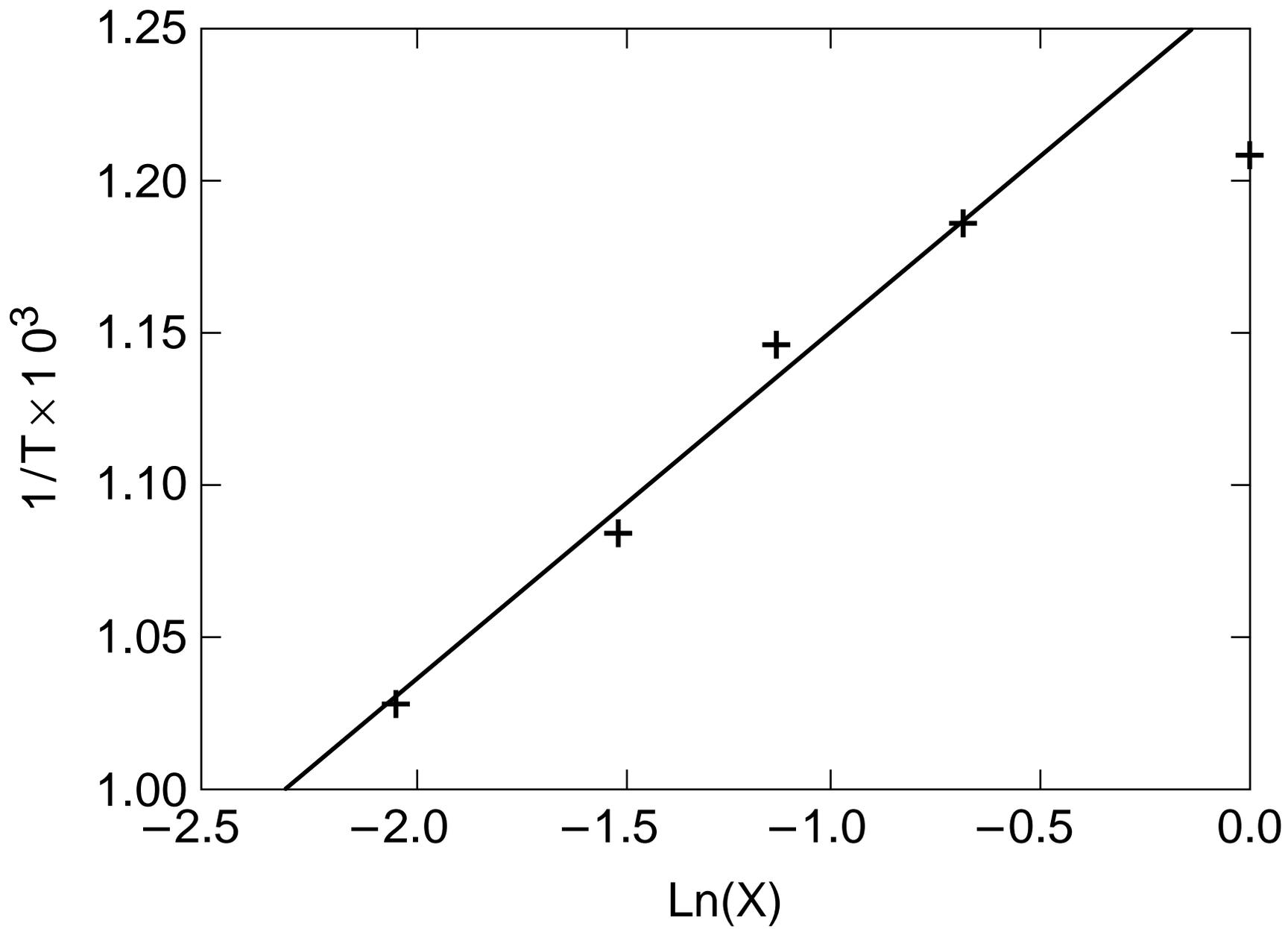
\end{document}